\documentclass[sigconf,nonacm]{acmart}
\AtBeginDocument{%
  }

\usepackage{array}[=2016-10-06]
\usepackage{algorithm}
\usepackage{algorithmic}
\usepackage{array}
\usepackage{booktabs}

\begin{document}

\title[Printegrated Circuits]{Printegrated Circuits: Personal Fabrication of 3D Printed Devices with Embedded PCBs}

\author{Oliver Child}
\affiliation{%
 \institution{University of Bristol}
 \city{Bristol}
 \country{UK}}
\email{oc17584@bristol.ac.uk}

\author{Ollie Hanton}
\affiliation{%
 \institution{University of Bath}
 \city{Bath}
 \country{UK}}
\email{oph33@bath.ac.uk}

\author{Jack Dawson}
\affiliation{
 \institution{University of Bath}
 \city{Bath}
 \country{UK}}
\email{jd2704@bath.ac.uk}

\author{Steve Hodges}
\affiliation{%
 \institution{University of Lancaster}
 \city{Lancaster}
 \country{UK}}
\email{steve.hodges@lancaster.ac.uk}

\author{Mike Fraser}
\affiliation{%
 \institution{University of Bristol}
 \city{Bristol}
 \country{UK}}
\email{mike.fraser@bristol.ac.uk}

\renewcommand{\shortauthors}{Child et al.}

\begin{abstract}

Consumer-level multi-material 3D printing with conductive thermoplastics enables fabrication of interactive elements for bespoke tangible devices. However, large feature sizes, high resistance materials, and limitations of printable control circuitry mean that deployable devices cannot be printed without post-print assembly steps. To address these challenges, we present Printegrated Circuits, a technique that uses traditional electronics as material to 3D print self-contained interactive objects. Embedded PCBs are placed into recesses during a pause in the print, and through a process we term \textit{Prinjection},  conductive filament is injected into their plated-through holes. This automatically creates reliable electrical and mechanical contact, eliminating the need for manual wiring or bespoke connectors. We describe the custom machine code generation that supports our approach, and characterise its electrical and mechanical properties. With our 6 demonstrations, we highlight how the Printegrated Circuits process fits into existing design and prototyping workflows as well as informs future research agendas.
     
\end{abstract}

\begin{teaserfigure}
  \includegraphics[width=\textwidth]{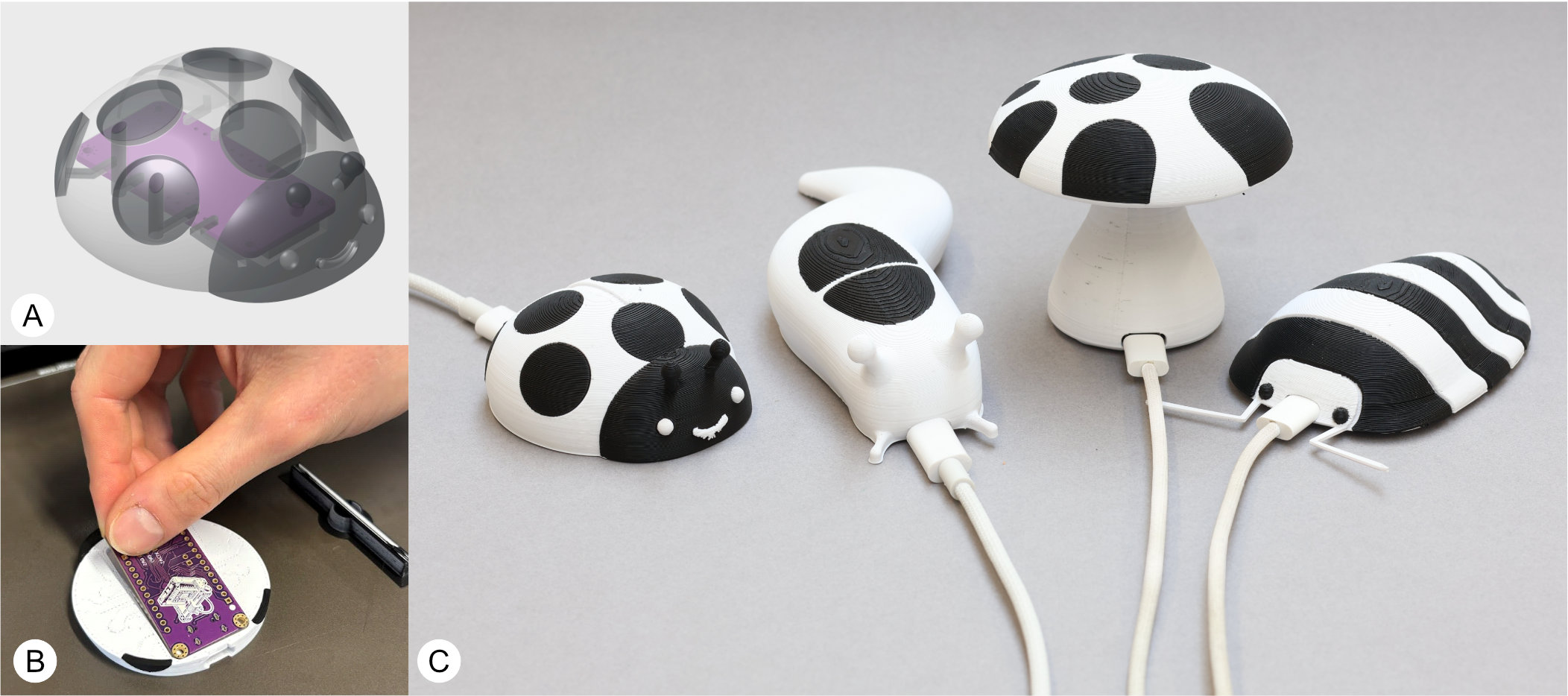}
  \caption{\textit{Printegrated Circuits} is a new approach that leverages multi-material 3D printers to create seamless interactive objects with embedded circuitry. (A) Interactive objects are designed with conductive elements and traces with cavities for embedded PCBs (B). During fabrication, the 3D printing process is paused to allow PCBs to be inserted inside an object. (C) Four example Printegrated Circuit designs that detect touch interactions and communicate over USB.}
  \Description{\textit{Printegrated Circuits} is a new approach that leverages multi-material 3D printers to create seamless interactive objects with embedded circuitry. (A) Interactive objects are designed with conductive elements and traces with cavities for embedded PCBs (B). During fabrication, the 3D printing process is paused to allow PCBs to be inserted inside an object. (C) Four example Printegrated Circuit designs that detect touch interactions and communicate over USB.}
  \label{fig:teaser}
\end{teaserfigure}


\maketitle

\vspace{-0.2cm}
\section{Introduction}
3D printers are now ubiquitous and are frequently used by makers, designers, students and researchers. Breakthroughs in print speed, price, and reliability of fused filament fabrication (FFF) 3D printers have led to much greater consumer adoption and applications~\cite{context_3d_printing_2024}. Increasingly common multi-material printers now allow practitioners to go beyond making parts of a single homogeneous material and combine thermoplastics of different colours and material properties in the same print. This unlocks a body of work exploring combining standard and conductive thermoplastics~\cite{leigh_simple_2012, flowers_3d_2017}. This research includes the creation of free-form capacitive touch sensors~\cite{brito_multimodal_2016, burstyn_printput_2015, schmitz_oh_2021}, extensions of existing touch interfaces~\cite{kato_3d_2016, marky_3d-auth_2020, ikematsu_ohmic-touch_2018, kato_capath_2020}, deformation sensing~\cite{sakura_lattisense_2023, schmitz_flexibles_2017}, and electric field sensing~\cite{schmitz_trilaterate_2019, iyer_3d_2017, zhang_electrick_2017, alalawi_mechsense_2023, gong_metasense_2021}, as well printed passive and active circuitry~\cite{canada_monolithically_2023, canada_semiconductor-free_2024, canada_three-dimensional_2024} and displays~\cite{hanton_protospray_2020}.

Echoing the revolution in desktop publishing enabled by the personal computer, rumblings of a revolution in "desktop manufacturing"~\cite{gershenfeld_fab_2005} enabled by digital fabrication tools are now louder than ever. Access to these tools rewrites the rules of production of digital devices, suggesting a way to unlock the long tail of hardware manufacturing~\cite{hodges_long_2019}, supporting the fabrication of customised products suited for individual needs.

In and beyond HCI, the goal of many of these developments is to support the fabrication of high-fidelity research prototypes~\cite{odom_research_2016} or customised interactive objects for a market of one~\cite{gershenfeld_how_2012}. Current demonstrations of 3D printed electronic interactive devices are still limited by the resistance of materials and the resolution of available machines~\cite{song_3d_2017}. Printegrated Circuits presents an approach that harnesses existing consumer FFF 3D printers, electronics, and materials to create self-contained printed prototypes. The key to Printegrated Circuits is confining high-resolution and resistance-dependent circuitry to traditional electronic material in the form of PCBs, and using the printer itself to automatically form connections between these PCBs and the free-form and customisable object.

The Printegrated Circuits process comprises two steps: 1) the automated pausing of the 3D printing process on a multi-material 3D printer, to allow manual insertion of a PCB-mounted microcontroller into a pre-designed cavity; and 2) the novel \textit{Prinjection} step we have developed, which extrudes conductive filament directly into plated-through holes (PTHs) of the inserted PCB. This creates robust electrical and mechanical connections to the microcontroller before the print continues.

To implement the Printegrated Circuits process, we have developed a slicer post-processing plug-in to allow users to specify the layer and position within the 3D printing process for PCBs to be inserted. The script splices the machine instruction file with commands to pause the print job and subsequently issue Prinjection movements to connect the 3D print and the PCB. The Prinjected filament also allows subsequent layers to form stable electrical and mechanical connections with the PCB. Through a series of described tests, we optimise and characterise the Prinjection movements, on boards that simulate inserted PCBs that would traditionally be used with breadboards, or soldered into projects. The process produces enclosed free-form devices with microcontrollers robustly fixed and connected within the device itself. The process is designed to be accessible to existing users of 3D printing technology through its easy integration into existing tool flows and use of tools already in the hands of makers. The process also supports recovery of embedded circuits to allow for iterative design and sustainable re-use of valuable components.

Our 6 demonstrations show how the technique works with different previously presented printed input modalities; custom, as well as off-the-shelf printed circuit boards; and haptic centralised applications. These demonstrations also provide the grounds for our discussion, where we highlight the opportunities and challenges in printing high-fidelity interactive prototypes, expressing the craft-like dialogue between the creative designer and the precise and repeatable machine.
\section{Related Work}

\subsection{High Fidelity Prototyping with Electronics as Material}

Personal fabrication in HCI supports democratised tangible device creation with a growing body of work incorporating electronics as material in craft-like prototyping. For instance, Little Bits \cite{bdeir_electronics_2009} presented the "electronics as material" abstraction to facilitate non-expert crafting of interactive devices with magnetic electronic modules that could be incorporated in all sorts of projects. Circuit Stickers \cite{qi_crafting_2015} took this further by re-packaging electronics as adhesive stickers to seamlessly integrate with hand-drawn paper-based circuits. Supported by the diverse ecosystem of prototyping kits and development platforms~\cite{lambrichts_survey_2021}, electronics are now integral to any prototyper's workflow at every level.

Paired with accessible electronics, digital fabrication machines have allowed individuals to create robust and appealing enclosures for crafted electronics~\cite{savage_makers_2015, weichel_enclosed_2013, shemelya_encapsulated_2015, arroyos_tale_2022}. However, increasing interest in supporting scalability and low-volume production of devices~\cite{boucher_research_2023, hodges_long_2019, khurana_beyond_2020} has drawn focus to the manual nature of this kind of making and the reliability needed for deployment. One approach to overcome this challenge is to directly harness the intrinsic scalability and reliability of outsourced production. Hartley et al. explore supporting the transition between prototyping module-based electronics and assembled PCBs in MakeDevice~\cite{hartley_makedevice_2023}. Others have gone further and used the PCB substrate itself as material to support the fabrication of complex and high fidelity devices~\cite{gonzalez_constraint-driven_2023, gonzalez_layer_2022, karras_pop-up_2017, dementyev_sensortape_2015}. 

This work addresses a key tension in interactive device creation: the trade-off between outsourced manufacturing and local fabrication. While outsourced production offers reliability and scalability, it lacks the tangibility, rapid iteration, and deep customisability inherent to hands-on, local making. Printegrated Circuits aims to resolve this tension by harnessing the best of both worlds, using commercially manufactured PCBs as material for rapid customisation with local digital fabrication tools.

\subsection{3D Printed Electronics for Interactive Devices}
3D printing technologies are expanding beyond the creation and replication of mechanical components to encompass rapid fabrication of electronics too. Commercial electronics printing techniques include multi-layered inkjet printing with conductive materials~\cite{botfactory_botfactory_2025, dragonfly_dragonfly_2025}, direct ink writing with silver paste~\cite{voltera_voltera_nodate, noauthor_httpswwwvoxel8com_2019}, multi-jet fusion with conductive powders~\cite{he_modelec_2022},  and aerosol jet systems~\cite{optomec_optomec_nodate}. The development of these approaches suggests a future where high-fidelity interactive devices can be rapidly prototyped, produced and customised to the needs of any individual. However, these specialised electronics printers remain out of reach of many designers, researchers, and makers, with machines requiring trained operators, being prohibitively expensive, or only existing in research and development settings, and not on the maker's desk~\cite{gershenfeld_fab_2005}.

In response to this, and towards the goal of democratising the use of digital machines to fabricate interactive devices, researchers have hacked and modified existing fabrication devices~\cite{nisser_laserfactory_2021, kawahara_instant_2013, yan_fibercuit_2022, imai_hot_2019, gao_revomaker_2015} and built alternatives~\cite{ziervogel_expansion_2021, peng_3d_2016, pourjafarian_print--sketch_2022, vasquez_jubilee_2020, sun_magnedot_2024, oh_pep_2018}. Others have augmented 3D prints made on low-cost machines to make resulting parts interactive through post-print augmentation~\cite{savage_series_2014, tseng_circuit_2021, bacher_defsense_2016, zhu_curveboards_2020, hook_making_2014, zhu_morphsensor_2020, palma_capacitive_2024, umetani_surfcuit_2017}, or mid-print part insertion~\cite{dauden_roquet_3d_2016, shemelya_encapsulated_2015}.

Material developments in printable thermoplastics have enabled researchers, designers, and makers to harness increasingly prevalent and reliable multi-material fused filament fabrication (FFF) 3D printers to print with carbon composite rigid thermoplastics~\cite{protopasta_3d_nodate, leigh_simple_2012, swaminathan_fiberwire_2019}, conductive TPU~\cite{sakura_lattisense_2023}, and copper-based materials~\cite{flowers_3d_2017}. Multi-material devices demonstrate functionality that can be incorporated into interactive devices such as capacitive touch sensors~\cite{brito_multimodal_2016, burstyn_printput_2015, schmitz_oh_2021}, deformation sensors~\cite{sakura_lattisense_2023, schmitz_flexibles_2017}, and position sensing through electric fields~\cite{schmitz_trilaterate_2019, iyer_3d_2017, zhang_electrick_2017, alalawi_mechsense_2023, gong_metasense_2021}.

FFF printing of functional logic or driver circuitry to pair with these interactive structures, while currently being investigated~\cite{canada_semiconductor-free_2024}, is limited by materials and resolution. Printed conductive structures are therefore commonly paired with traditional electronics. This has been done by wiring directly between print and electronic circuit by melting wires into the surface~\cite{protopasta_3d_nodate}, using silver paste~\cite{swaminathan_fiberwire_2019}, or copper plating combined with press-fit connections~\cite{hong_thermoformed_2021, li_e-joint_2024}, introducing technical manual steps to the making process. Oh Snap!~\cite{schmitz_oh_2021} overcomes this challenge with a custom piece of electronics to interface between the PCB and print, with additional magnetic elements and pogo pins. Thermopixels~\cite{moon_3d_2024} prints directly over plated pads with vias and uses the same circuitry used to thermally activate the pixels to force high current through the print in order to heat the filament and ensure sufficient electrical connection to the metal.

These approaches support the reliability of connections through the use of custom-designed elements in the connected circuitry. The Printegrated Circuits approach uses an un-modified 3D printer, accessible tools and can combine them with traditional prototyping electronics hardware through the use of custom automated print paths.

\subsection{Custom 3D Printer Workflows for Interactive Devices}

Going beyond the fabrication of homogeneous plastic parts requires custom design and fabrication workflows and explorable parameter spaces~\cite{subbaraman_its_2025, twigg-smith_tools_2021} unconstrained by abstractions that traditionally encapsulate domain-specific knowledge~\cite{fossdal_vespidae_2023}.

Bespoke CAD tools support the design of specific elements that allow for interactivity within 3D prints. In \textit{A Series of Tubes}~\cite{savage_series_2014}, Savage et al. demonstrate a tool for routing pipes through 3D models to add interactivity through conductive pathways and air-filled pipes. X-Strings~\cite{li_xstrings_2025} is a tool to design and print string-actuated mechanisms straight off the printer. \textit{Capricate}~\cite{schmitz_capricate_2015} presents a design tool to add capacitive touch pads and automatically place conductive traces.

Some workflow modifications involve mid-print part-insertion-based interventions~\cite{wall_intermittent_2025}.  For example, to reduce print time and wasted material, Substiports~\cite{wall_substiports_2023} and Scrappy~\cite {wall_scrappy_2021} use existing objects for support and internal media. Medley~\cite{chen_medley_2018} and Encore~\cite{chen_encore_2015} combine existing materials and objects into 3D prints to augment their function. To make traditionally static 3D printed objects interactive, Folded Printgami~\cite{dauden_roquet_3d_2016} inserts folded paper-based circuits mid-print. Rhapso creates functional string-based mechanisms~\cite{ashbrook_rhapso_2024} and uses a custom thread-placing mechanism to integrate new material properties into prints. Liquidio~\cite {schmitz_liquido_2016} prints cavities and manually inserts liquid to allow for different tilt sensing modalities by creating contacts between thermoplastic electrodes. 

Creating tools to generate domain-specific tool-paths for FFF printers is also an interesting avenue of research, allowing researchers to think of 3D prints as more than homogeneous parts, and imparting qualities through printing that can't be described in the design stage. Specifically, Wasserfall et al.~\cite{wasserfall_topology-aware_2020}  suggest a way to integrate wire generation into a slicing program, and Duty et al.~\cite{duty_z-pinning_2019} suggest injecting molten filament into infill gaps to increase the vertical strength of parts.

Printegrated Circuits adds to this body of work by suggesting mid-print intervention and domain-specific tool-path generation: PCBs are inserted during a pause in printing, and custom tool-paths we call "Prinjection" allow for conductive filament to be vertically injected into the plated-through holes of the PCB. This workflow enables the fabrication of fully self-contained interactive devices.

\section{Printegrated Circuits}
Drawing from the related work, Printegrated Circuits presents a solution to locally fabricate high-fidelity prototypes using electronics as material, using an unmodified multi-material 3D printer with conductive materials and custom tool-paths.

\begin{figure}
    \centering
    \includegraphics[width=.85\linewidth]{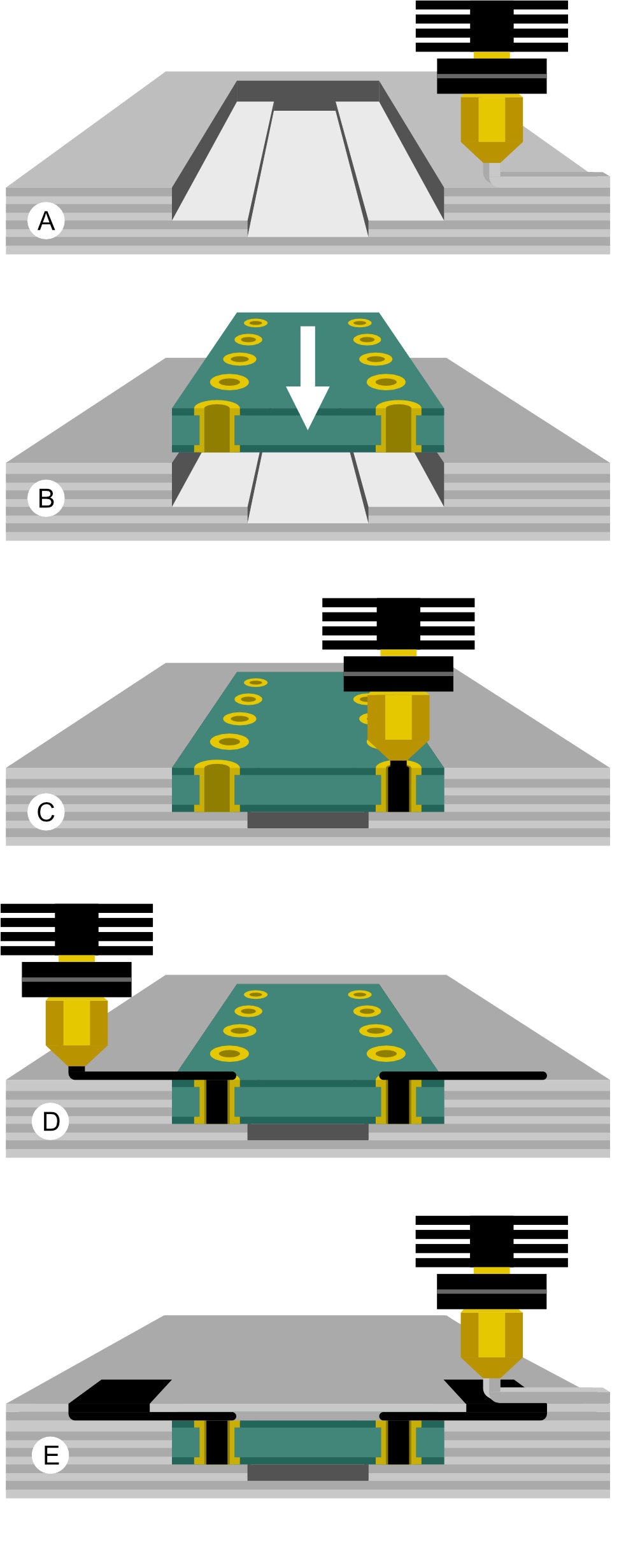}
    \caption{The Printegrated Circuits process (A) The object is printed with non-conductive material following commands generated by slicing software. (B) The printer pauses for manual insertion of the PCB. (C) The printer injects a fixed amount of conductive filament directly into plated-through holes in the PCB, in a process we term \textit{Prinjection}. (D) Conductive traces are printed from the Prinjected contacts. (E) The rest of the object and its interactive structures are completed.}
    \label{fig:printegration}
    \Description{A vertical reel showing the timeline of the Printegrated Circuits process. Each of the 5 pictures shows a different stage of a 3D printed part having the board embedded in it. The first picture shows layers of a 3D print with a cutout and a 3D printer nozzle extruding white plastic. The second shows a green PCB being placed in the cutout. The third has the 3D printer head come back and is extruding black material into the holes on the PCB. The fourth, shows the printer extruding more black plastic traces over the PCB and existing whit part. The final image shows more layers of black and white filament being built up.}
\end{figure}

\begin{figure*}[ht!]
    \centering
    \includegraphics[width=1\linewidth]{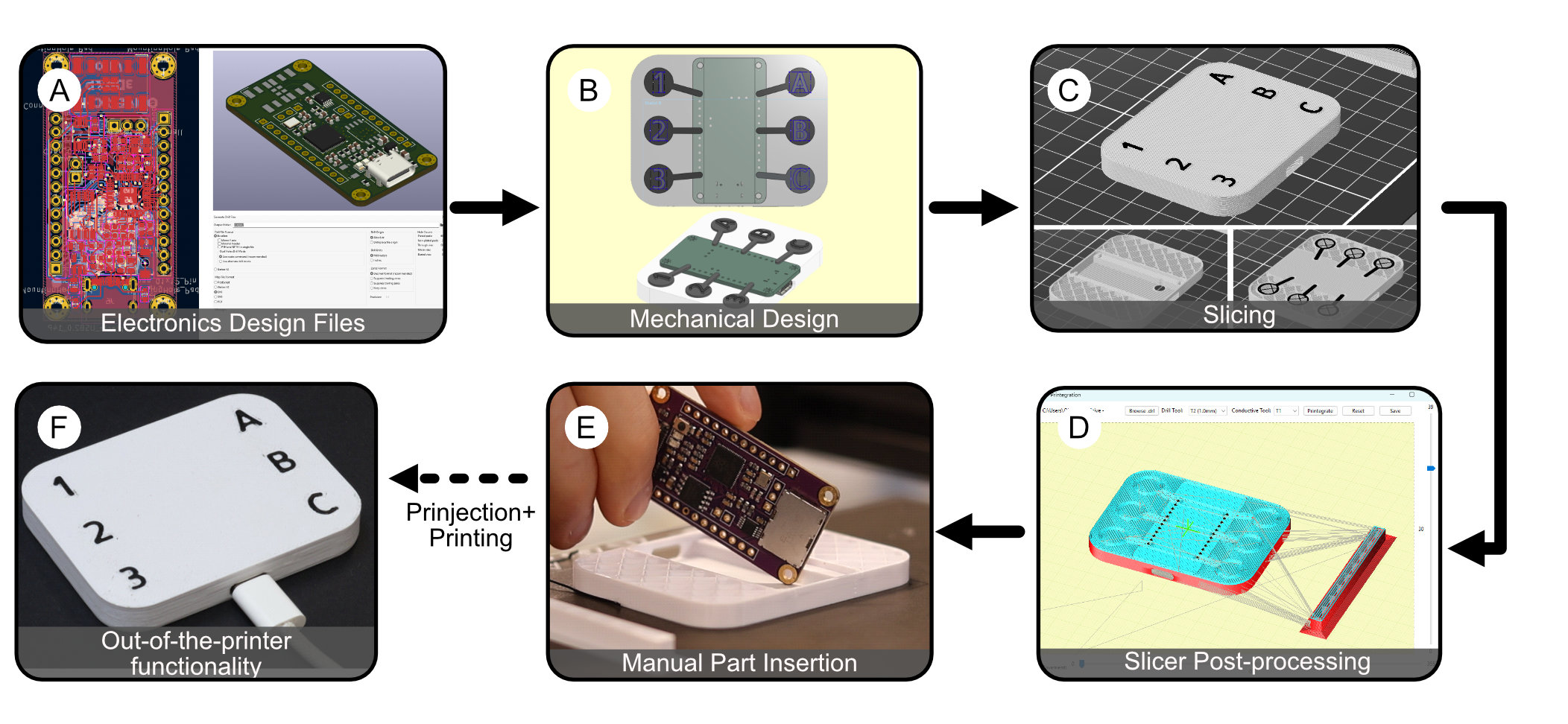}
    \caption{The Printegrated Circuit workflow: (A) PCB designs in EDA exported as STEP with .drl file. (B) In-context CAD modelling of an interactive object with embedded PCB (C) The slicing step (D) The sliced file is re-interpreted in a post-processing script, merging in instructions to support the necessary Prinjection steps (E) Designs are printed with a mid-print pause for part insertion followed by Prinjection (F) Printegrated Circuits work out-of-the-printer.}
    \label{fig:workflow}
    \Description{A picture representing the workflow of Printegrated Circuits composed of 6 sub assemblies of pictures. A shows the PCB design step. On the far left a PCB design file is show with red and blue traces displaying where copper is on the top and bottom of the board. Along side this is a rendering of the PCB, it's a rectangular board with rounded corners and holes along both sides. Below this is a computer dialogue showing the export stage in the software. The next step shows two images of a 3D modelled device. It has 6 buttons, each with traces leading to the PCB designed in the previous step. One of the images is a rendering of the transparent device face down, while the other is an isometric view. The next stage is represented by 3 Images of the slicing process. The same model displayed in the previous step is then shown in a different software where the layer lines that the 3D print would produce can be seen. There are two images showing different slices of the print. One where there is only white material, and the next here the buttons are starting to be printed. The final step is represented by another 3 images. One shows a computational notebook with a single slice of the print plotted in a graphing tool, The remaining two images are photographs of the device, one during printing where a hand is coming in to place the board, and another where it is completed and plugged in with a USB cable}
\end{figure*}

As illustrated in Figure \ref{fig:printegration}, the Printegrated Circuits fabrication process involves: pausing the print for parts to be manually inserted; injecting conductive filament into the plated-through holes normally used for soldering, a process which we call Prinjection; and connecting conductive sensing structures with the PCB.

Printegrated Circuits use manually designed recesses and conductive traces modelled within the objects to seamlessly accommodate integration of conventional electronics that are inserted during the print process. The Prinjection process is automatic, with custom g-code movements spliced into the print file with a slicer post-processing script directing the printer to form robust mechanical and electrical connections to the PCB and creating anchors for further layering of traces with conductive filament. 

Printegrated Circuits dovetails with existing device prototyping paradigms, allowing for the insertion of off-the-shelf and custom-designed PCBs to be used in the process. These PCBs are treated as reusable materials, where plastic can be softened with a heat gun and high-value modular electronics recovered for future use in design iterations, or altogether new devices.

\subsection{The Printegrated Circuits Workflow}


In Figure \ref{fig:workflow}, we present each workflow stage for an example scenario creating an illustrative "ABC 123" USB capacitive touch HID keyboard device with arbitrary characters using the Printegrated Circuit Board introduced in Section \ref{printegrated circuit board}. Here, we give a high-level overview of the most important details of the process. In our supplementary material, we provide in-depth details of specific constraints and design recommendations. 

\subsubsection{Electronic CAD}
The workflow uses the digital design files of the embedded PCB. Therefore, the first stage is to export a STEP model (.step) of the PCB and drill file (.drl) to represent the positions of the plated-through holes. These can be exported from electronic CAD platforms, or open-source designs can typically be downloaded directly from the supplier of prototyping PCBs.

\subsubsection{Mechanical CAD}
Printegrated Circuits are modelled around recesses for the embedded PCBs, which are designed using the .step file exported from ECAD. PCBs must be placed parallel to the printing surface with the unpopulated face pointing upwards. Sensing structures and traces are also manually designed at this stage. This design is realised within parametric timeline-based CAD programs like Onshape\footnote{Onshape, \url{https://onshape.com/}, Accessed 2025-03} and Autodesk Fusion\footnote{Autodesk Fusion, \url{https://www.autodesk.com/ca-en/products/fusion-360}, Accessed 2025-03}. The process is not dependent on any specific CAD platform, giving practitioners autonomy over the design of novel features. Conductive elements are modelled as separate parts and exported from the CAD platforms as a 3MF file.

\subsubsection{Slicing (CAM)}
We perform slicing using Prusa Slicer V2.7.4\footnote{Prusa Slicer, \url{https://www.prusa3d.com/page/prusaslicer_424/}, Accessed: 2025-03}. Each part of the model is assigned either a conductive or non-conductive material at this stage. This is a standard feature of slicing software, normally used to set the colour of different parts of a model. Printing infill is set to 100\% for conductive parts to increase conductivity, and internal interface layers between parts are enabled to ensure high fidelity of the internal structure. We include a slicer configuration file with all of the settings in our supplementary material.

\subsubsection{Printegrated Circuits Post-processing Script}
We use the post-processing script setting in Prusa Slicer. On g-code export of print files, the slicer runs a script that opens our graphical program based on Pronterface\footnote{Pronterface, \url{https://www.pronterface.com/}, Accessed: 2025-03}  allowing the user to specify the position and layer of the to-be-inserted PCB and its associated plated-through holes on a representation of the sliced print. After selecting the board and placing the holes, the user clicks the `Printegrate' button, and g-code lines are inserted into the print file specifying the pause, tool changes, extra nozzle priming movements, and the Prinjection steps. We describe the specific stages of Prinjection in more detail in Section \ref{prinjection}.

\subsubsection{Printing (CNC)}
Our examples and demonstrations were built with a Prusa XL 5T tool-changing FFF 3D printer with 0.4mm nozzle diameters\footnote{Prusa XL 5T, \url{https://www.prusa3d.com/en/product/original-prusa-xl-semi-assembled-5-toolhead-3d-printer/}, Accessed: 2025-03}. This printer is supplied with 5 print heads at around 4000 USD, targeting 3D printing hobbyists as well as professionals.

We used Protopasta Conductive Composite PLA (90 USD/kg)\footnote{Protopasta Electrically Conductive Composite PLA \url{https://proto-pasta.com/products/conductive-pla}, Accessed: 2025-03}, for conductive traces and white eSUN PLA+ (20 USD/kg)\footnote{ eSUN PLA +, \url{https://www.esun3d.com/pla-pro-product/}, Accessed: 2025-03} as the structural material, but other non-conductive thermoplastics would work equally well.

In striving to make the Printegrated Circuits process more adoptable by the maker and research community, a discussion of the practicality of using other machines and materials is included in the supplementary materials of this paper.

\subsection{Prinjection}\label{prinjection}

\begin{figure*}
    \centering
    \includegraphics[width=\linewidth]{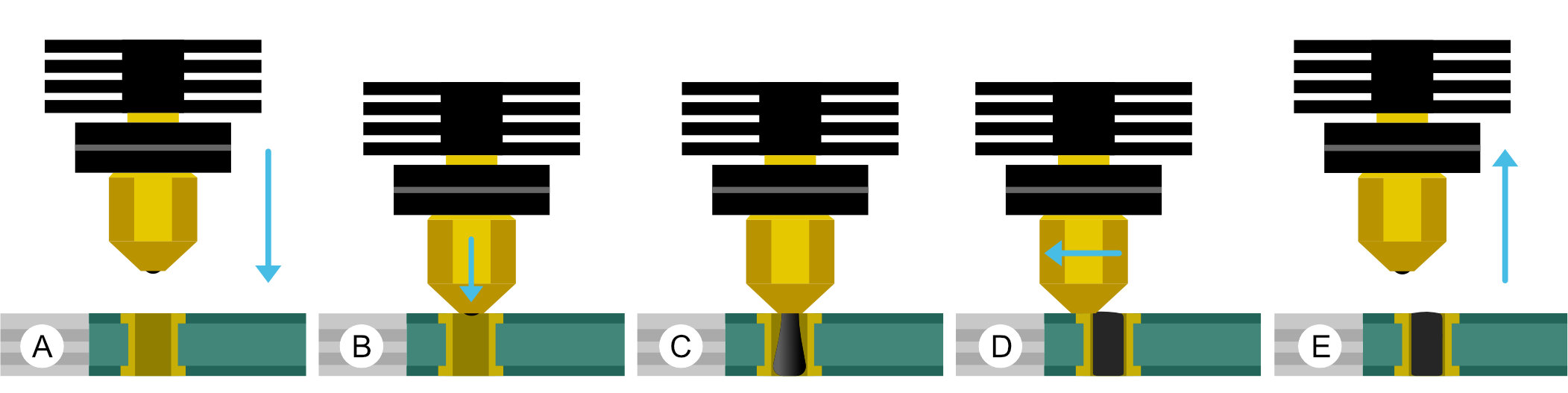}
    \caption{Prinjection Movements: After priming the nozzle, (A) the tool moves over each plated-through hole, (B) lowers onto the surface, (C) injects conductive plastic into the through hole, (D) wipes the nozzle against the surface of the PCB, before (E) lifting the tool back up.}
    \label{fig:Prinjection}
    \Description{Prinjection moves displayed graphically in 5 Steps: Each vector image shows a 3D printer head, a PCB and associated 3D print. In the first image, the printer head is a small distance above the PCB, and an arrow indicates it is moving downward. In the next image the nozzle is right on top of the PCB. In the third image there is now conductive material emerging from the nozzle into a through hole on the PCB. In the fourth, an arrow shows that the nozzle is moving to the left, leaving the black extruded material in the hole. Finally, in the last image, the nozzle moves back up.}
\end{figure*}

The Prinjection technique uses custom-generated g-code to inject conductive filament into the plated-through holes of an embedded PCB. PCBs are embedded during a user-specified pause in the printing process, and Prinjection locations are calculated using the user-specified PCB insertion position and the plated-through hole coordinates retrieved from the PCB's original manufacturing or footprint files. Prinjection uses the pressure of the injected molten plastic to conform to the edges of the surrounding metal, similar to injection moulding. The large contact area reduces contact resistance and improves the mechanical reliability of the formed connections. Prinjection also creates a mechanical anchor for conductive traces, overcoming the challenges faced by low adhesion to the PCB's solder mask. 

Figure \ref{fig:Prinjection} demonstrates a simplified version of the individual Prinjection steps, each movement controlled by a series of g-code instructions written below.

\begin{algorithm}
\caption*{Prinjection Commands} 
\begin{algorithmic}
\FOR{each hole in PTHs}
    \STATE $\text{gcode} \mathrel{+}= \text{"G0 X"} + \text{hole.x\_coord} + \text{" Y"} + \text{hole.y\_coord}$
    \STATE $\text{gcode} \mathrel{+}= \text{"G0 Z"} + \text{board\_height}$
    \STATE $\text{gcode} \mathrel{+}= \text{"G1 E"} + \text{(retract\_amount + extrude\_amount)}$
    \STATE $\text{gcode} \mathrel{+}= \text{"G1 E"} + \text{(-retract\_amount)}$
    \STATE $\text{gcode} \mathrel{+}= \text{"G0 X"} + \text{(hole.x\_coord + sweep)}$
    \STATE $\text{gcode} \mathrel{+}= \text{"G0 Z"} + \text{(board\_height + lift)}$
\ENDFOR
\end{algorithmic}
\end{algorithm}

\begin{figure}
    \centering
    \includegraphics[width=\linewidth]{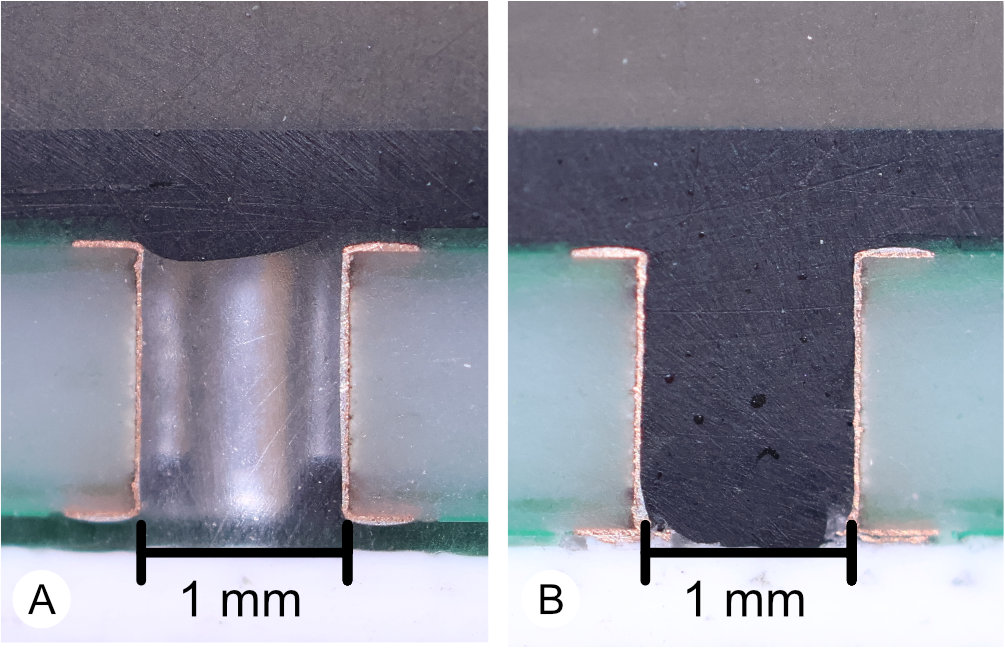}
    \caption{Cross-sections of the plated-through holes of a test PCB cast in resin, sanded and polished. (A) Conductive filament is printed onto the PCB surface without Prinjection; it lies on the surface making little contact with metal. (B) Prinjection leaves holes fully filled and ensures reliable connectivity.}
    \Description{Cross-sections of the plated-through holes of a test PCB cast in resin, sanded and polished. (A) Conductive filament is printed onto the PCB surface without Prinjection; it lies on the surface making little contact with metal. (B) Prinjection leaves holes fully filled, creating good electrical contact and a mechanical anchor for subsequent layers.}
    \label{fig:cross-section}
\end{figure}

Figure \ref{fig:cross-section} shows the cross-section of a successful Prinjection of a plated-through hole in a prototyping board against a non-Prinjected sample, where both samples have had a conductive trace laid over the top of the through hole. The samples were potted in resin, cut, and repeatedly sanded and polished following the process described by Schlaepfer and Oskay~\cite{oskay_open_2022}.
\\ 

\begin{figure}
    \centering
    \includegraphics[width=\linewidth]{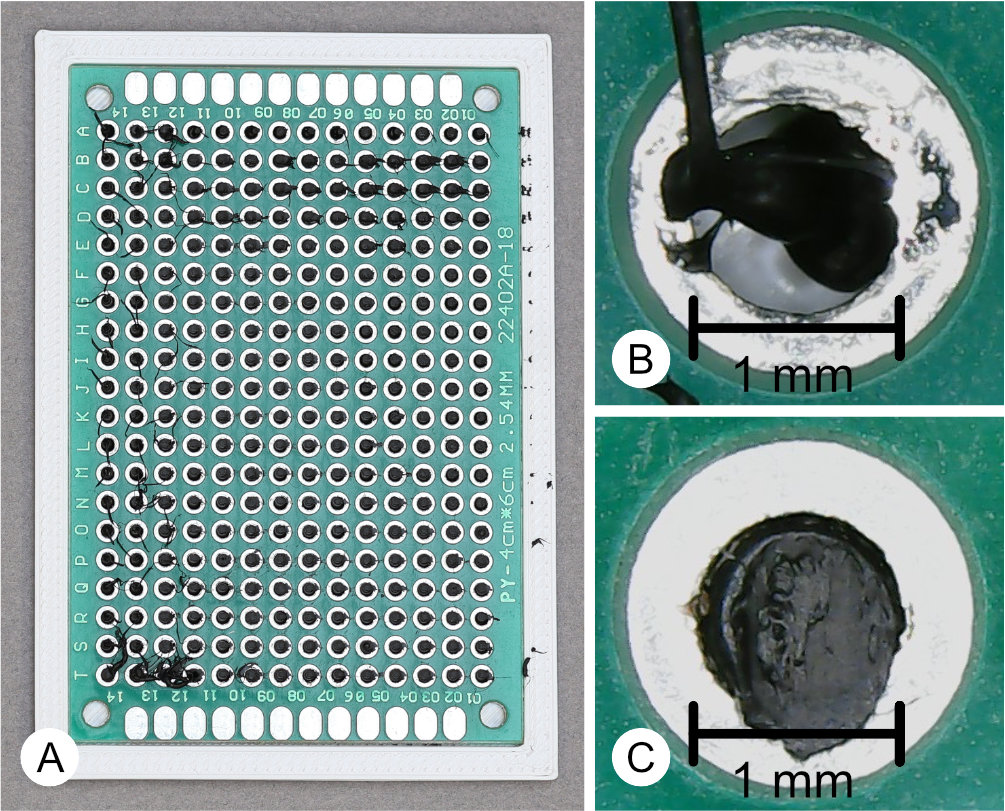}
    \caption{Prinjection optimisation tests using a prototyping board. (A) The holes in a column are filled using identical Prinjection commands, before making a single parameter change for the next column. (B) An under-filled plated-through hole compared with (C) a perfectly-filled hole.}
    \label{fig:hole filling}
    \Description{3 Pictures. The first shows a whole "proto-board" a green PCB with a rectangular array of holes each filled with black plastic. The whole board is surrounded by a white plastic 3D printed frame. The other two pictures are close-ups of a single hole. B shows a small amount of black plastic in the hole, and you can see the white frame underneath. C shows a fully filled hole with a slight amount of wiped material just below on the surface of the plated ring. }
\end{figure}

\begin{figure*}[h]
    \centering
    \includegraphics[width=1\linewidth]{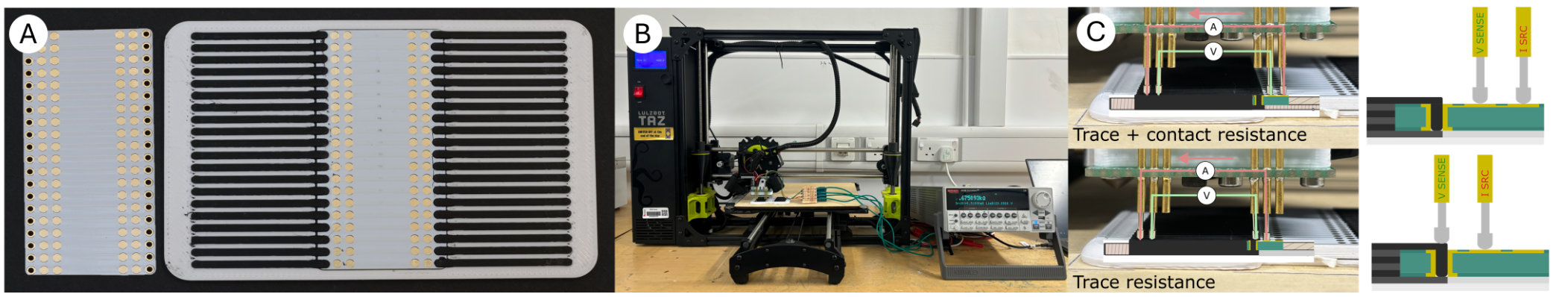}
    \caption{Contact resistance characterisation: (A) Custom PCB with contacts for 4-point probe. (B) Automated testing of contact resistance using an upcycled Lulzbot 3D printer, (C) Two different configurations of pogo pins in a custom probe for 4 4-point probe.}
    \label{fig:characterisation printegrated circuit}
    \Description{(A)     displays the white characterisation PCB before and after Prinjection. Before Prinjection you can see through the plated-through holes as well as see the gold pads. Following the process, black traces are shown over the where the holes were, it's all encased in a white frame. In the next picture (B) there is a 3D Printer frame that is being used as an automated stage hooked up to a piece of test equipment. One of the circuits is lying on the print bed being probed. the last image shows the different circuits being created by the custom probe when it is being tested.}
    
\end{figure*}

\section{Characterisation and Calibration of Prinjected Contacts}
The goal of these characterisations and tests was to optimise and show how prinjection could be used to make reliable electrical and mechanical contact between 3D printed object and metal on an embedded PCB. The tests used PCBs with 2.54 mm or 1/10" spaced 1mm diameter plated-through holes, as this is what is common in prototyping settings where boards are traditionally inserted in breadboards.

\subsection{Optimising Prinjections}
We identified a set of optimal printer movements and parameters for the injection of conductive material into plated-through holes on a PCB to form the base of electrical connections to the rest of a 3D printed object. The key goals were to adequately fill the holes with conductive material so that there was surface contact on all sides and to provide a surface for the next planar layers of conductive material to adhere to without too much excess extruded material or strings that could cause unwanted paths between neighbouring contact points.

To explore optimal parameters, we used blank prototyping boards\footnote{Universal prototyping boards, \url{https://shop.pimoroni.com/products/universal-proto-board-pcbs-3-pac}, Accessed 2025-04} with 14 columns of 20 plated-through holes typical of a breakout board, coupled with a 3D printed frame to hold the prototyping board in place, as shown in Figure \ref{fig:hole filling}A. Samples were prepared to test one parameter at a time. After placing the board in the frame, we instructed the 3D printer to perform Prinjection steps on each of the through holes with conductive filament. For each column, the printer would first perform a wipe movement that would prime the nozzle with conductive material, then perform Prinjection movements for each plated-through hole in that 20 hole column. This was repeated for each column of holes with the same parameter incremented by a set amount for each column. The parameters we varied were: extrusion amount, retraction amount, and wipe length. The same movement with the same values was repeated for a whole column of 20 holes because we wanted to understand whether changes in pressure in the nozzle were occurring. After each test, we inspected the samples and visually identified which holes were most uniformly filled with minimum stringing like the example in Figure \ref{fig:hole filling}C.

\subsubsection*{Extrusion length}
Our first test determined the correct extrusion volume. We tested extrusion amounts from 0.2mm to 0.9mm of standard 1.75mm diameter filament in increments of 0.05mm. We selected this range because the theoretical total volume of a 1.0mm diameter cylindrical plated-through hole in a standard 1.6mm thick PCB is 1.6mm * $\pi$ * 0.5mm$^2$ = 1.26mm$^2$. This volume constitutes a length of around 0.74mm of 1.75mm diameter filament. For each extrusion length, a column of 20 consecutive holes was filled. After each row, the printer performed a wipe move to re-prime the nozzle ready for the next column of holes.

An extrusion length of 0.65mm gave the best results, just below the theoretical through hole volume. However, significant stringing was still apparent in all tests, including when using the optimal length. Our next test, therefore, focused on reducing stringing.


\subsubsection*{Retraction}
The second test explored retraction of the filament, with the goal of reducing the stringing that was apparent in the previous test. Pulling back filament before moving the nozzle can help reduce stringing.

While we found that stringing is not often a catastrophic issue in Printegrated Circuits - subsequent print head passes when printing conductive traces will normally fragment and remove the strings - but there is still a small risk that residual filament between adjacent holes could cause  unintended crosstalk or other signal integrity issues, if not a complete short circuit.

We performed retraction tests from 1mm to 8mm in increments of 0.5mm. Higher retractions resulted in less stringing. However, we did not exceed 8mm of retraction because above this value the nozzle became clogged. Too much retraction also decreases the overall print duration and causes temperature creep as heat travels back up the extruder. When plastic passes its glass transition temperature and cools too far up the nozzle, it can cause the filament to jam. After visually inspecting the results, we determined a retraction amount of 7.5mm gave the least stringing without negative side-effects.

\subsubsection*{Wiping}
The third test explored nozzle wiping. After each injection, we dragged the nozzle at the same z-height off to the side of the hole. This was to try and `cut off' any strings of filament and produce a cleaner pattern.

Following a similar approach to the above optimizations, we filled in a grid of plated-through holes, gradually increasing the wipe length. A longer wipe length proved more reliable, but we balanced this against increased print duration and the risk that, for a more realistic PCB, long wipes might interfere with nearby components and features. From our tests we determined a wipe length of 2mm was enough for the nozzle to completely clear the plated-through hole.

\subsection{Contact Resistance} \label{contact resistance}

We also explored the contact resistance of the Prinjected material. Contact resistance is a measure of the additional impedance introduced by the interface between adjacent conducting materials. Poor adhesion, low contact surface area and unstable mechanical connections can all increase contact resistance, limiting the functionality of fabricated devices.


We calculated the contact resistance at the PCB-to-PLA interface by measuring the difference between the resistance across the length of a conductive trace and the resistance of the length of the trace connected to a plated-through hole. Direct resistance measurement with a multimeter is inaccurate as it varies with the pressure applied to the probes, which themselves introduce a variable contact resistance. We obtained accurate measurements using the 4-point probe technique, eliminating the resistance of the probes and the contact between the probe and material. While the resistance of the probes themselves was relatively low, the contact resistance between the probe and material was highly variable.

We took measurements with a Keithley 2614B source measurement unit (SMU) connected to a bespoke 4-point probe assembly, which we designed and built. The probe used 4 dome-capped pogo pins mounted in a linear arrangement. We also developed a custom contact resistance characterisation PCB shown in Figure \ref{fig:characterisation printegrated circuit}A. The characterisation PCB comprised 20 traces with a plated-through hole, and two surface pads at either end. We designed the PCB with the same pitch as the pogo pins, so that from the end of each trace one pad could be used for the low force pin, the other for the low sense, and the through hole for the Prinjection process. The high sense and high force pins were then connected via the pogo pins at the other end of the characterisation board as shown in Figure \ref{fig:characterisation printegrated circuit}C. To ensure repeatable and consistent sensing, a re-purposed 3D printer stage was used to lower the probe assembly to a set height for each trace as shown in Figure \ref{fig:characterisation printegrated circuit}B. The modified 3D printer combined with the SMU could then take resistance measurements for a whole row of traces automatically. Changing the position of the force and sense pins on the PCB side of the probe by swapping pogo pins between connections supported a comparison between total resistance (trace + contact) and trace resistance alone as shown in \ref{fig:characterisation printegrated circuit}C. Both types of measurement were taken for each trace, and then the contact resistance was calculated by taking the difference between the two readings.

Two samples were prepared with Protopasta's conductive PLA with 3 mm deep, 25 mm long traces with a width of 1.5 mm. One sample was sliced with default slicer settings, a 0.2mm layer height and a pause command inserted through the slicer's interface at layer 13 for PCB insertion. The other sample was sliced in the same way but then underwent splicing with Prinjection commands, with the optimally selected movement settings from the previous tests.

\begin{figure}
    \centering
    \includegraphics[width=\linewidth]{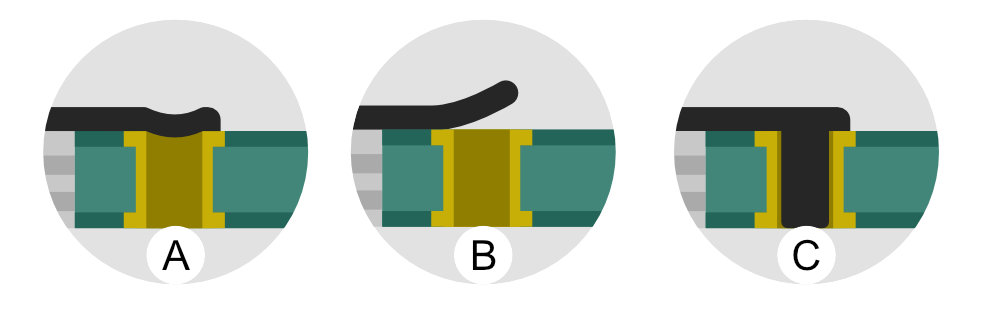}
    \caption{Without Prinjection (A, B), the printed conductive trace has nothing to anchor to, resulting in the printed trace peeling up from the surface. With Prinjection (C) printed traces become anchored and do not peel up.}
    \label{fig:peel}
    \Description{Two representations of inline-layers of plastic, with printed circuit and a 3D printer head extruding plastic, on the left, the trace is laid down onto the surface of the plastic and PCB, and the trace sinks down into the hole, on the right, the trace has lifted up above the surface.}
\end{figure}

\begin{figure}
    \centering
    \includegraphics[width=\linewidth]{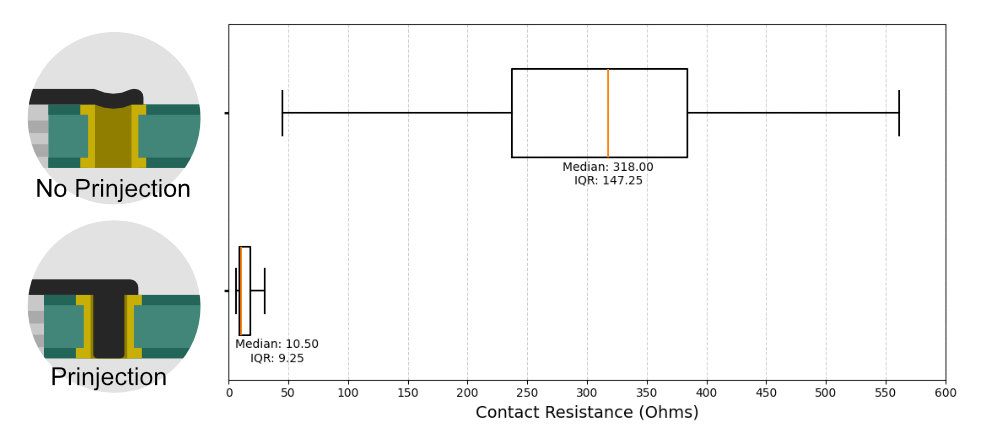}
    \caption{Characterisation of the contact resistance of a planar print vs. with Prinjection, see Section \ref{contact resistance} for details.}
    \label{fig:enter-label}
    \Description{Box plots displaying a much wider range of resistances for planar prints rather than prinjected prints. Next to each of the boxes is a graphical representation of the planar vs prinjected setup.}
\end{figure}

An independent samples one-tailed T-test revealed a significant difference in contact resistance distribution between the planar and Prinjection samples, t(74)=7.06, p<0.001. The Prinjection contact resistance was the same order of magnitude as the trace resistance of the material, while the resistance of the planar connection was significantly higher. Importantly, the Prinjection contact resistance also showed much lower variance, useful for ensuring predictable electrical properties at the modelling stage, and when optimising or calibrating a particular Printegrated Circuit design. In contrast, only 36 of the 40 planar traces formed an electrical connection (i.e. there was a 10\% failure rate). The smooth surface of the PCB and the lack of anchoring offered by a Prinjection plug meant that in the 4 failed cases, the overlaid conductive trace peeled up during printing, as depicted in Figure \ref{fig:peel}B. This peeling effectively resulted in an open circuit.

Our test validates the benefit of Prinjection over the planar layering which has been used in previous work to print conductive filament over PCBs, such as ThermoPixels~\cite{moon_3d_2024}. The greater mechanical reliability and consistently stable electrical properties allow repeatable production of Printegrated Circuits using the parameters described above.

\subsection{Robustness} \label{robustness}
\begin{figure}
    \centering
    \includegraphics[width=\linewidth]{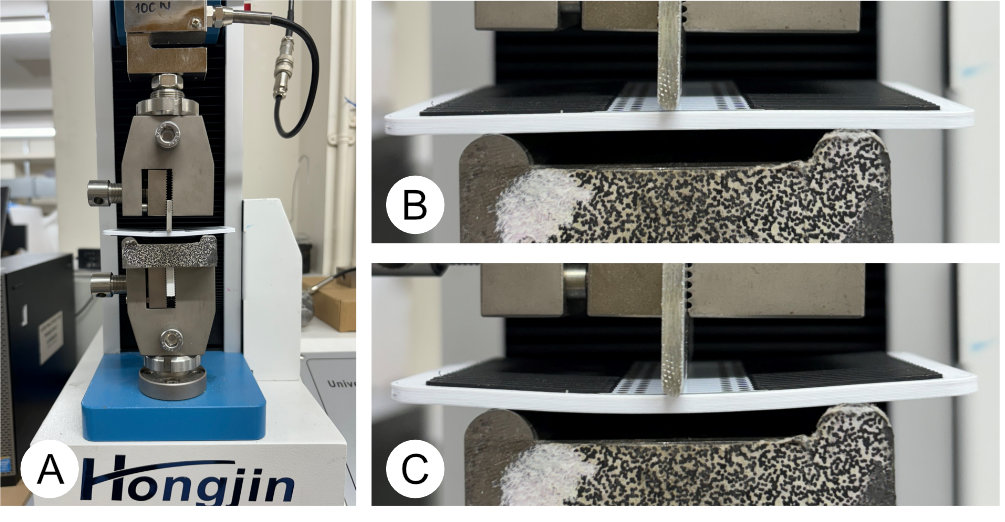}
    \caption{Robustness testing with a universal testing machine in a three-point bending setup (A) and applying 75 N of force repeatedly to the sample board (B-C)}
    \label{fig:universal testing}
    \Description[]{3 images of a universal testing machine. One is scaled out with clamps holding a bar above and two bars below in order to bend a prinjection sample. On the right there are two images displaying the sample in its bent and unbent state}
\end{figure}

Building on the consistency of our contact resistance measurements, we further explored the physical robustness of Printegrated Circuits. Robustness is a multifaceted concept; our approach was to further explore the quality of electrical contact by taking measurements before and after repeated stress testing, and comparing samples created using Prinjection with planar printing.

\subsubsection*{Method} Two samples were printed, one with planar printing and one with Prinjection. Both used contact resistance characterisation PCBs as in the previous test. A  Hongjin universal testing machine\footnote{Hongjin Universal Testing Machine, \url{https://www.hongjin-group.net/good-quality-universal-testing-machine-universal-tensile-testing-machine-hongjin-3-product/}, Last Accessed: 2025-04} equipped with a 3-point bend fixture and 100N strain gauge was used. For each sample, 100 bend cycles were performed with a 75N force. After 50 cycles the sample was turned over to perform the remaining 50 bend cycles from the other side. Contact resistance measurements were taken before and after these cycles.

\subsubsection*{Results}
An independent samples one-tailed T-test compared changes in contact resistance before and after bending across the Prinjected and planar samples. Our test showed a significant difference in contact resistance t(78)=-1.67, p<0.05). The test showed that the Prinjected sample was significantly more robust to bending than the planar sample.

We were surprised that, unlike the planar sample, the Prinjected sample actually showed a reduced average contact resistance following repeated bending, demonstrating remarkable electro-mechanical robustness. Further study is required, but it appears that mechanical manipulation actually improved surface-to-surface contact of the Prinjected material.

\subsection{Summary of Results}

Our tests provide optimal measures for the Prinjection technique, and demonstrate its importance in providing robust electrical and mechanical connections with significantly lower and more predictable resistance when interconnecting conventional and 3D printed electronics.

The evaluation work was conducted with a simple sample assembly which does not represent the breadth of design or fabrication space of the Printegrated Circuits approach. The characterisation board and experiments were optimised to identify fundamental characteristics of the Printegrated Circuits. In the following section, we go beyond test scenarios and show some example demonstrations that we have built to give a sense of the opportunities and potential that this robust 3D printed device workflow provides.

\begin{table*}[th]
\centering
\caption{Demonstrations of the Printegrated Circuits method with fabrication information}
\label{tab:3d_print_demos}
\begin{tabular}{@{}l>{\raggedright\arraybackslash}p{3cm}p{3.5cm}>{\raggedright\arraybackslash}p{2.5cm}p{0.9cm}p{1.2cm}p{1.2cm}@{}}
\toprule
Device & Function & Interaction Modalities & Embedded Circuit & PLA (g) & Conductive PLA (g) & Print Time \\ \midrule
Slug & Scrolling device/slide clicker & Cap Touch \& Vibrotactile & Printegrated Circuit Board & 32 & 6 & 2h15 \\
Ladybird & Media controller & Cap Touch \& Vibrotactile & Printegrated Circuit Board & 33 & 13 & 2h35 \\
Isopod & Stroke sensitive emoji keyboard & Cap Touch \& Vibrotactile & Printegrated Circuit Board & 29 & 7 & 2h39 \\ 
TuneShroom & USB MIDI Keyboard & Cap Touch & Seeed Xiao ESP32-S3 & 73 & 23 & 8h22\\ 
SnailSense & IOT Plant Pot Moisture Sensor & Resistance Sensing & Seeed Xiao ESP32-S3 & 31 & 25 & 3h59\\
Lego Data Physicaliser & Block Sensing Bar Graph & Resistive \& Contact & Seeed Xiao ESP32-S3 & 78 & 77 & 7h25\\
\bottomrule
\end{tabular}
\end{table*}

\section{Demonstrations and Properties of Printegrated Circuits}

In this section, we present six demonstrative devices we have built during the development of Printegrated Circuits. These demonstrations are shown in Figures \ref{fig:teaser}, \ref{fig:customisable}, \ref{fig:tuneshroom} and \ref{fig:snail}, and are summarised in Table \ref{tab:3d_print_demos}. Printegrated Circuit devices are inherently user-ready through seamless design, off-the-printbed functionality, freedom of form and no required post-assembly. Through specific demonstrations we provide examples of integration within personal fabrication ecosystems, enabling fully customisable device creation (\ref{customisable}), supporting iterative design (\ref{iterative}), integration within existing design processes (\ref{functional}), support for low volume production (\ref{scalable}) and compatibility with established fabrication workflows (\ref{existing workflows}). We further evidence Printegrated Circuit's inherent capacity to overcome material challenges (\ref{materialchallenges}), the robustness of fabricated devices (\ref{robust demo}) and their highly sensitive functionality (\ref{sensitive}). Through these demonstrations, we highlight the approach's advantages and aim to inspire further exploration by underlining opportunities and challenges.



\subsection{The Printegrated Circuit Board} \label{printegrated circuit board}

\begin{figure}[b]
    \centering
    \includegraphics[width=1\linewidth]{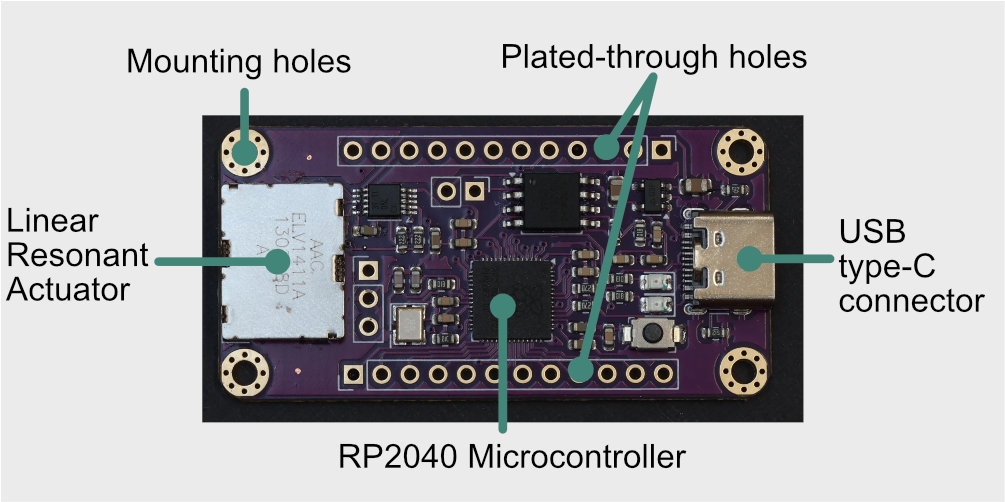}
    \caption{Our Printegrated circuit board combines the RP2040 microcontroller with a linear resonant actuator, allowing for capacitive touch sensing paired with centralised vibrotactile response. The Printegrated Circuits process can use this board or off-the-shelf prototyping boards. This}
    \label{fig:capsense haptic PCB}
    \Description{A purple printed circuit board with USB connector, pin holes and metal can that contains the Linear Resonant Actuator}
\end{figure}

An advantage of Printegrated Circuits is that they work with both existing prototyping PCBs such as those available from suppliers like Adafruit\footnote{Adafruit Industries, \url{https://www.adafruit.com/},  Accessed: 2025-03} and Seeed Studio\footnote{Seeed Studio, \url{https://www.seeedstudio.com/}, Accessed: 2025-03}, as well as custom platforms that can be designed using conventional electronics design tools and readily available components. 

However, to support our demonstrations, we developed our own open-source general-purpose PCB, which enables the fabrication of objects that combine capacitive touch and haptic feedback, shown in Figure \ref{fig:capsense haptic PCB}. This Printegrated Circuit Board combines an RP2040 microcontroller\footnote{RP2040 Microcontroller, \url{https://www.raspberrypi.com/products/rp2040/}, Accessed: 2025-03} and a Linear Resonant Actuator (LRA) to create haptic effects.

Capacitive touch sensing through loading mode self-capacitance~\cite{grosse-puppendahl_finding_2017} is implemented with a single pin with precisely timed capacitance measurements carried out using the microcontroller's Programmable IO (PIO). Using the microcontroller's pins directly means that they can be easily reconfigured for other output and sensing too.

The board uses a surface mount haptic actuator\footnote{ ELV1411a LRA, \url{https://nfpshop.com/product/lvm-series-linear-vibration-motors-elv-1411a}, Accessed 2025-03}, which is well suited to Printegrated Circuits because like the microcontroller they can be centralised and placed on a PCB embedded in the part and have actuation projected onto the surface. Single sources of vibrotactile stimulation can create a number of illusions~\cite{kildal_3d-press_2010}.

The example we have already shown in Figure \ref{fig:workflow} uses our Printegrated Circuit Board for a general-purpose application. In the demonstrations below, we show how it is re-used in more complex devices.

\subsection {Printegrated Circuits are Customisable} \label{customisable}

\begin{figure*}
    \centering
    \includegraphics[width=1\linewidth]{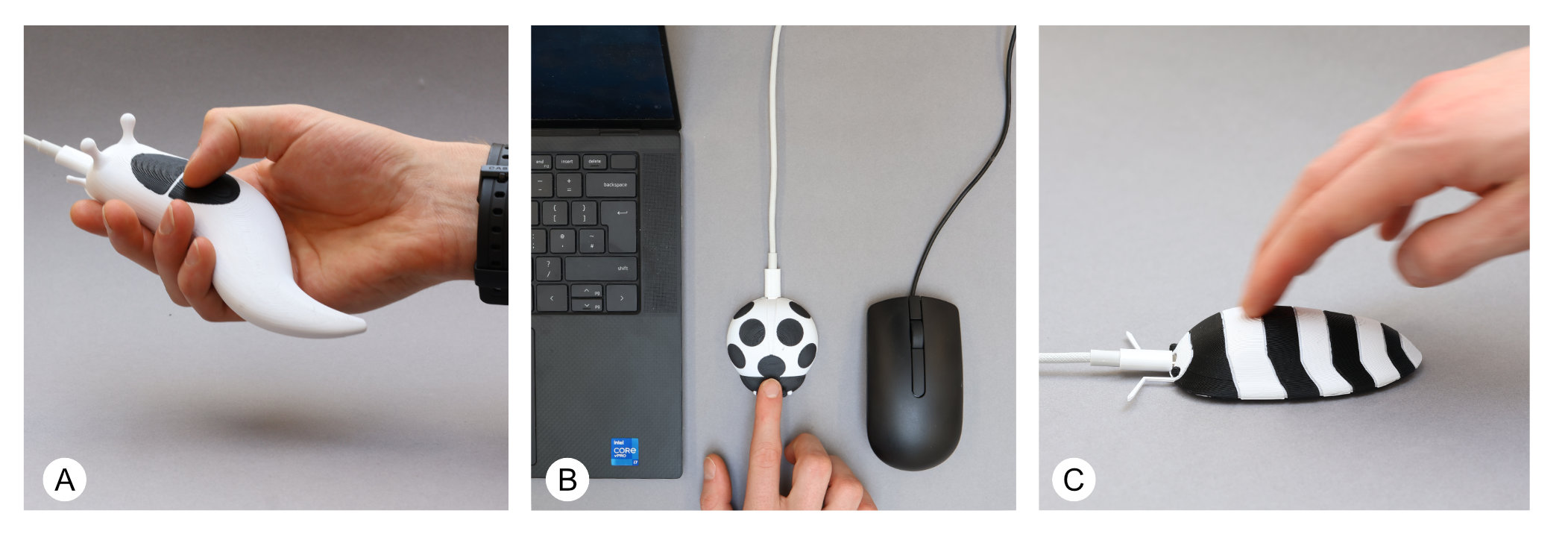}
    \caption{The slug, ladybird and isopod are all USB HID devices derived from the same base functional design, but each is customised to have a unique form and function. (A) the slug is a haptic presentation clicker, (B) the ladybird is a media controller, (C) the isopod is a 
    stroke-sensitive desk pet.}
    \label{fig:customisable}
    \Description{The slug, ladybird and isopod are all USB HID devices derived from the same base functional design, but each is customised to have a unique form and function. (A) the slug is a haptic presentation clicker, (B) the ladybird is a media controller, (C) the isopod is a stroke-sensitive desk pet.}
\end{figure*}

\textit{The IO-creatures} (Figure \ref{fig:customisable}) are a series of capacitive touch USB human input devices (HID) with haptic feedback. They all use the Printegrated Circuits board described earlier. The design of the devices stem from the same origin, the enclosed ABC123 board also presented in the earlier sections. Designers can customise devices for specific uses through digitally designed form, and firmware controlled function. The IO-creatures demonstrate how Printegrated Circuits prioritises customisation and freedom of form.

Specifically, each of the IO-creatures has a different behaviour or function. Using CircuitPython running on the Printegrated Circuit Board, different keystroke behaviours can be programmed when capacitive touch events are registered. These can be mapped to any of the pins on the Printegrated Circuit Board, from which channels to the conductive pads are printed.

\textbf{\textit{The Slug}}  (Figure \ref{fig:customisable}A) has two capacitive touch pads mapped to the up and down arrows so that it can be used as a presentation clicker, with the vibration providing non-visual feedback. 

\textbf{\textit{The Ladybird}} (Figure \ref{fig:customisable}B) behaves as a media controller with different capacitive spots controlling volume, play/pause, and skip forward/backwards. 

\textbf{\textit{The Isopod}}  (Figure \ref{fig:customisable}C) measures consecutive touches along its back to detect stroking. When stroked down its back it prints happy emojis, but when stroked backwards, it sends keystrokes encoding angry emojiis to the host machine. It also makes use of the haptic actuator to "purr" through vibration effects when stroked.

\subsection{Printegrated Circuits Support an Iterative Design Process}\label{iterative}

\begin{figure*}[t]
    \centering
    \includegraphics[width=1\linewidth]{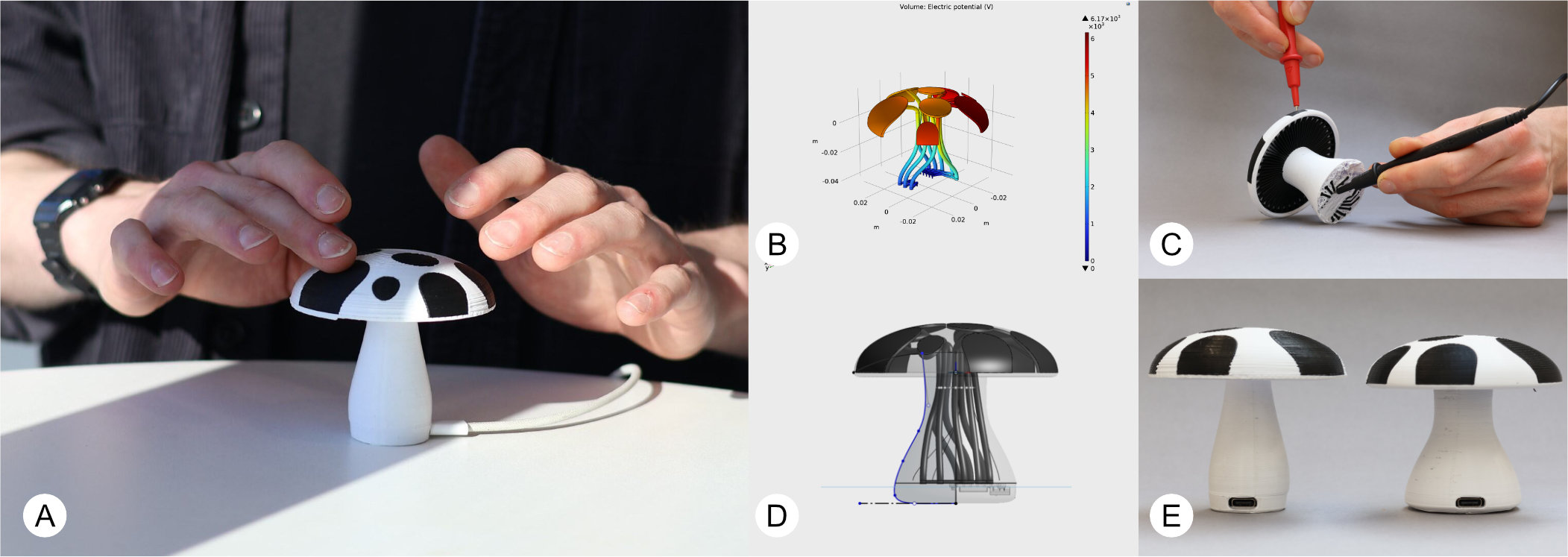}
    \caption{The TuneShroom is a MIDI instrument. Its digital design means that it can be both simulated and tested in the physical world with the ability to make rapid modifications to the design.}
    \Description{The TuneShroom is a MIDI instrument. Its digital design means that it can be both simulated and tested in the physical world with the ability to make rapid modifications to the design.}
    \label{fig:tuneshroom}
\end{figure*}

\textbf{\textit{The TuneShroom}}  (Figure \ref{fig:tuneshroom}) is a mushroom-shaped MIDI instrument. Finger touches on the black spots of the cap are registered through capacitive touch sensing. The embedded board, a Seeed Studio Xiao ESP32-S3\footnote{Seeed Studio Xiao ESP32 S3, \url{https://www.seeedstudio.com/XIAO-ESP32S3-p-5627.html}, Accessed 2025-03}, uses USB MIDI to send notes to a host device such as a smartphone or computer running a MIDI synthesiser application or digital audio workstation (DAW).

The TuneShroom is an example of a more complex design, with challenging 3D routing between the base of the stem, where the microcontroller is inserted, and the pads on the cap of the mushroom. Printegrated Circuits supports iterative processes that are required when designing complex objects, harnessing local digital fabrication practices rather than outsourced production.

Like the other demonstrations, the Tuneshroom was designed in the browser-based CAD platform, Onshape. The platform uses a parametric, timeline-based workflow which allows for modification of previous features, supporting this iterative design workflow where early versions or sections of the design can be printed to inform design choices and changes.

While other researchers have presented tools for 3D routing specifically for the integration of interactive elements into 3D prints~\cite{savage_series_2014}, we chose to route our models by hand. This decision was made for three primary reasons: When testing existing solutions, they weren't able to correctly resolve the complex geometries we were working with more than two traces; routing by hand allows for the application of domain-specific knowledge to the specific scenario, e.g. understanding the impact of layer heights and anisotropic properties of materials and; finally automatic routing would constitute an independent step outside of the parametric design tools the rest of the model was designed in, which means that manual designs had the benefit that small design changes do not require re-performing any routing.

The physical design of the TuneShroom took several iterations to get right. A specific consideration was the design of the gills. We wanted to use the natural droop of unsupported molten filament to create organic-looking curved gills. This involved printing the device a number of times with manually designed supports. As a designer, having direct access to 3D printing tools meant that several iterations could be made in the time that it would take to have one version outsourced.

With this approach to making interactive devices, iteration is just as easy at all design stages. After the initial creation stage, and some use of our original design, we discovered that the TuneShroom had a flaw: the small base meant that when placed on a surface and a touch pad at the edge was tapped to play a note, it was unstable and prone to falling. The timeline-based CAD model made it easy to edit one of the original shape-defining sketches to stretch out the base and re-print the device with a wider base, as shown in Figure \ref{fig:tuneshroom}.

\subsection{Printegrated Circuits Enable Integration within Existing Design Processes}\label{functional}

\begin{figure*}[t]
    \centering
    \includegraphics[width=1\linewidth]{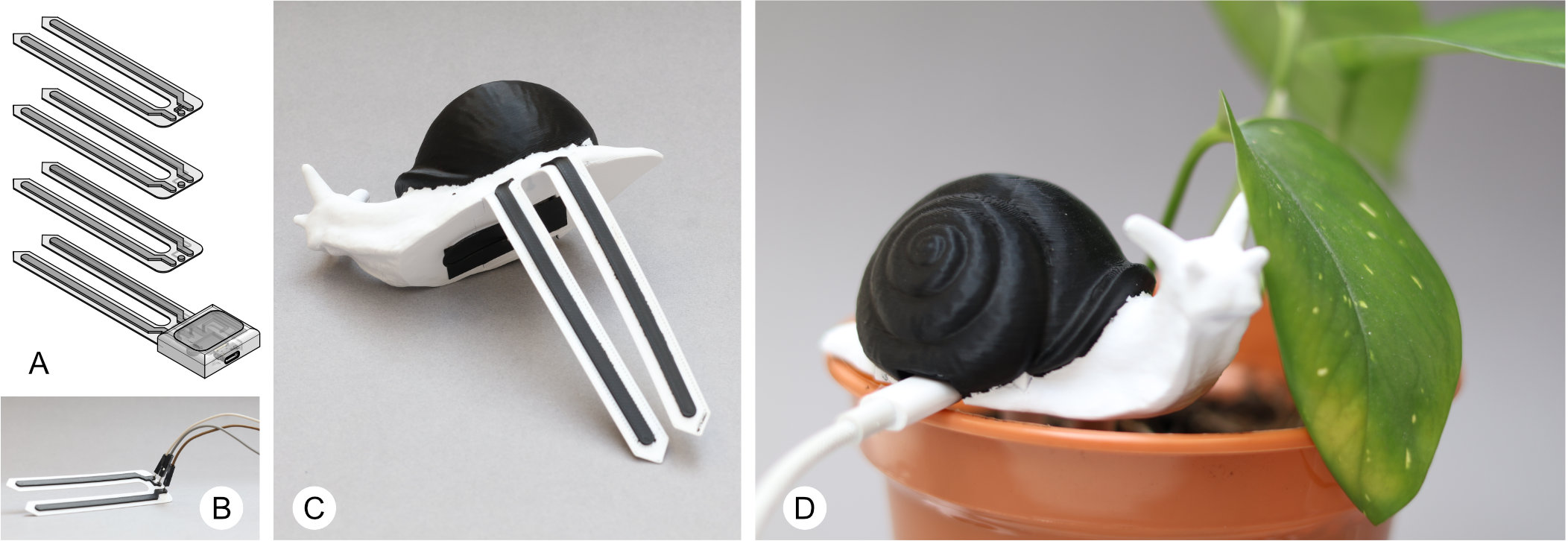}
    \caption{SnailSense is a plant pot moisture sensor that uses the resistance of the soil to track watering. It demonstrates how Printegrated Circuits allow iterative design through prototyping of sensing elements (A) and (B) before integration in a fully Printegrated 3D form (C) and (D). }
    \label{fig:snail}
    \Description{SnailSense is a plant pot moisture sensor that uses the resistance of the soil to track watering. It demonstrates how Printegrated Circuits allow iterative design through prototyping of sensing elements (A) and (B) before integration in a fully Printegrated 3D form (C) and (D). }
\end{figure*}

\textbf{\textit{SnailSense}} (Figure \ref{fig:snail}) is a plant pot moisture sensor based on Thingiverse user 3Dmon's snail model\footnote{3DMON Snail Thingiverse, \url{https://3d-mon.com/}, CC BY-SA, Accessed: 2025-03}. It uses resistance sensing to measure soil moisture. Dry soil has a high resistance and, when soaked, the resistance decreases. In contrast with the capacitive touch sensing devices, this device uses a voltage divider circuit that is designed into the object for sensing, as opposed to simple one-pin electrodes. The soil acts as a resistor between reference ground (0V) and an analogue pin on the microcontroller, a thin winding trace is designed into the print between sense and the high level (3V3) pin. The winding trace is designed to be comparable to the resistance of the soil, so the voltage read on the analogue pin is proportional to the resistance of the soil.

Before designing and printing a full object with embedded circuitry, iterative steps using existing prototyping tools like Arduinos and jumper cables can be performed. For example, in order to optimise the resistive sensing of the snail, different electrode designs shown in Figures \ref{fig:snail} A and B were first printed, tested and modified as stand-alone parts before combining as one in the snail. Specifically, we varied the length of the fixed resistor trace that had to be in a range similar to the resistance of the soil to maximise sensitivity. In the test pieces, the electrodes were manually connected through wires to an Arduino. Since robustness and minimising assembly time were not critical at this stage, making a full printegrated circuit wasn't necessary. Once optimised, though, these features could all be integrated into a single design, allowing for a fully integrated and repeatable print. 

\subsection{Printegrated Circuits Enable Fabrication of Modular and Scalable Devices}\label{scalable}

\begin{figure}[b]
    \centering
    \includegraphics[width=1\linewidth]{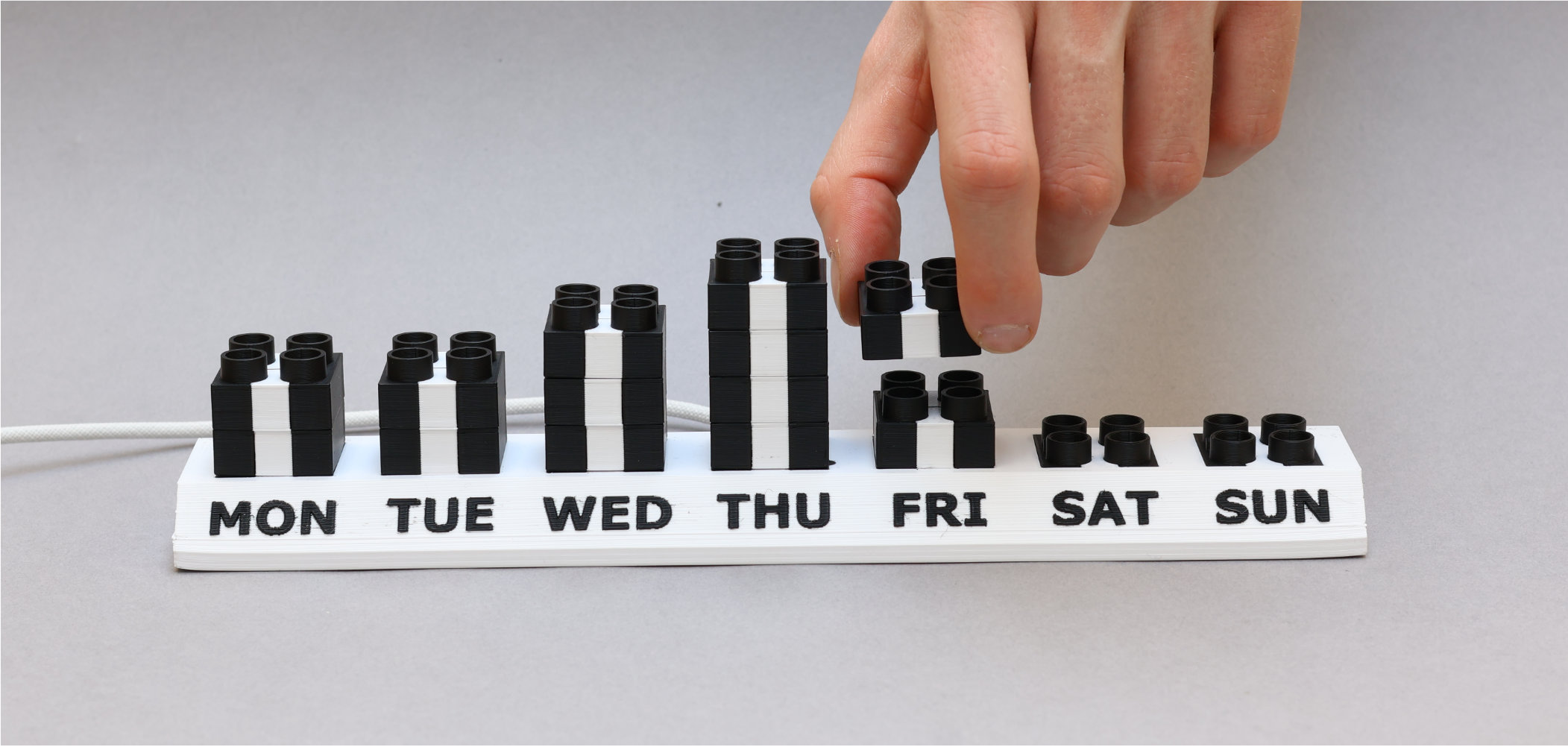}
    \caption{The Data Physicalisation Device uses the resistance of Lego-inspired towers to calculate height, allowing users to record information through tangible media.}
    \Description{The Data Physicalisation Device uses the resistance of Lego-inspired towers to calculate height, allowing users to record information through tangible media.}
    \label{fig:lego}
\end{figure}

\textbf{\textit{The Lego Data Physicaliser}} (Figure \ref{fig:lego}) is a personal habit tracker, designed to record the number of daily achievements through a tangible medium. The circuit measures the changes in resistance of each tower as bricks are added. The tower of bricks adds resistors in parallel to the circuit, which can be sensed through an analogue pin on the microcontroller. Daily counts are then recorded, allowing for seamless interactions between tangible control and digital tracking. Modular tangible interactive devices are an attractive avenue for researchers exploring the intersection of physical and digital, offering novel ways to explore data~\cite{chan_capstones_2012,yoshida_capacitive_2015}. One of the challenges in getting these devices into users' hands is the difficulty of fabricating many of the same modules. We present this device to show how passive electronic elements can be scaled to make duplicates and extensions to modular sensing systems without extra effort. In the context of personal fabrication for prototyping, we suggest a key benefit of Printegrated Circuits is that it enables simple scaling-up in the production of tangible interactive devices for user studies.

\subsection{Printegrated Circuits Integrate within Existing Workflows} \label{existing workflows}

When printing the SnailSense, the electrodes are printed flat on the printbed, however, for their application shown in Figure \ref{fig:snail}, they need to be pointing downwards to be inserted into the soil. In order to achieve this, we heated the relevant point in the model with a heat gun and folded it using an approach similar to Thermoformed Circuit Boards~\cite{hong_thermoformed_2021}. Unlike that prior work, however, the high resistance of the traces in our application is not a challenge, so no copper plating step is required. In fact, the high resistance allows for the design of passive resistors into the model that give rise to the voltage divider network needed for this application. This demonstrates that Printegrated circuits can harness the advantages of existing workflows to enable the fabrication of free-form devices.

\subsection{Printegrated Circuits Overcome Material Challenges}\label{materialchallenges}

The high resistance of carbon-black-based conductive filaments can be measured physically as well as simulated. In Comsol Multiphysics\footnote{Comsol Multiphysics, \url{https://www.comsol.com/}, Accessed: 2025-08} we can use the 3D geometry of the TuneShroom's traces and the material's anisotropic resistances to perform Finite Element Analysis to calculate the resistance of the traces.

The physics simulation was set up to calculate the resistance between the contact points on the PCB and the surface of the touch pads on the TuneShroom's cap. The conductive material's anisotropic resistance, caused by the 3D printer's layering of material, is used in the simulation by defining its conductivity with a diagonal matrix. We performed these simulations using the earlier measured resistance values.
$$
\sigma =
\begin{bmatrix}
5.55 & 0 & 0 \\
0 & 5.55 & 0 \\
0 & 0 & 3.33
\end{bmatrix} \quad \text{(S/m)}
$$

Figure \ref{fig:tuneshroom} (B) shows the voltage potential through the geometry of the TuneShroom's traces. The simulation results showed that the different conductive geometries all had a resistance between 4.5-6.1 k$\Omega$. These were compared with the directly measured resistances, which were all higher than the simulations (mean +1.1  k$\Omega$). This could be due to variations in print line orientations. This high resistance, however, doesn't pose an issue for the capacitive touch measurement as the low-resistance dependent control circuitry is all confined to the conventional PCB.

\subsection{Printegrated Circuits are Robust} \label{robust demo}
While the robustness of prinjected contacts was demonstrated through rigorous tests in section \ref{robustness}, the TuneShroom's durability has been evidenced through its use at multiple in-person events, where it was carried loosely in a backpack or pocket without any protective enclosure. Its robustness was further illustrated in an informal drop demonstration, where it was dropped from a height of around 2 meters ten times at various angles. Despite the repeated impacts, all touch pads remained fully functional, showing that the internal connections held up well.

\subsection{Printegrated Circuits are Sensitive} \label{sensitive}
The ESP32-S3 found in the Xiao microcontroller board used in the TuneShroom has native touch peripherals that can take 16-bit  capacitance readings. The design guidelines\footnote{ESP32 Touch Sensor Application Note, \url{https://github.com/espressif/esp-iot-solution/blob/release/v1.0/documents/touch_pad_solution/touch_sensor_design_en.md}, Accessed: 2025-03} recommend using an insulating cover over electrodes; to avoid extra coating steps and increase sensitivity with high resistance traces (recommended max of 2 k$\Omega$), we just use the bare conductive filament. This results in high capacitance readings when the pads are touched, up to the maximum readable value. Using the calculation described in the application note, where the sensitivity is $(\text{non-trigger value} - \text{trigger value}) / \text{non-trigger value} \times 100\%$ we measured the average sensitivity of 110\%, for our electrodes, despite the high resistance of the traces. To detect touch events, capacitance readings overcome a threshold. We found that this does not need to be calibrated per-pad or across devices.

\subsection{Recovery and Recycling}

\begin{figure*}
    \centering
    \includegraphics[width=\linewidth]{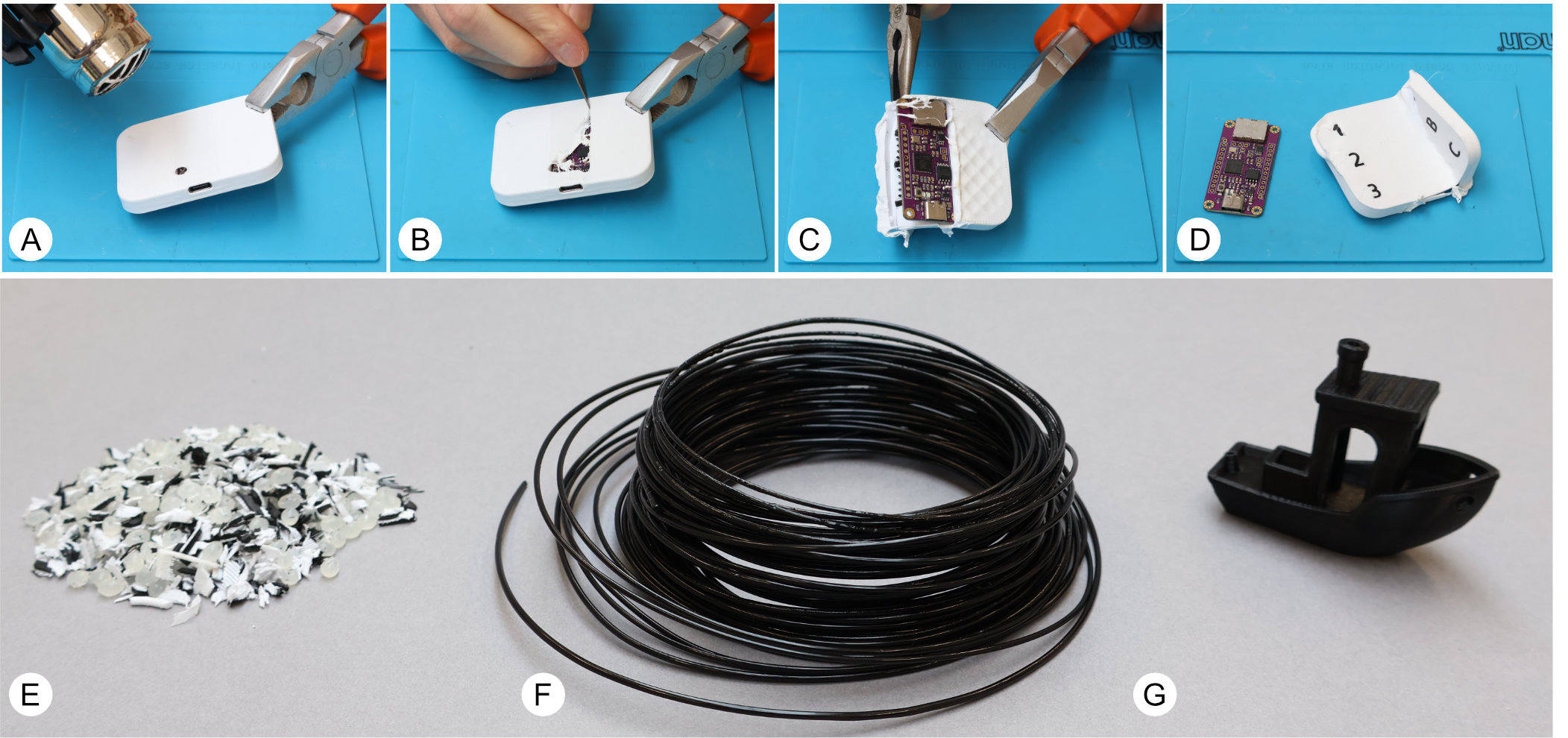}
    \caption{The process of recovering PCBs and filament from Printegrated Circuits: heating with a heat gun (A), using a scalpel to cut layers (B), using pliers to bend material when malleable (C) and freeing the circuit (D). Shredded parts are mixed with Virgin PLA pellets at a 33:67 ratio (E) to extrude new filament (F) that can be used to print new objects like this Benchy (G).}
    \label{fig:deconstructionl}
    \Description{Left to right: pliers holding a white rectangle and a heat gun pointed at the rectangle, pliers holding the same white rectangle as a hand uses a pencil knife to cut around an PCB embedded in the rectangle, pliers holding the same white rectangle as the hand now uses a different pair of pliers to remove plastic around the PCB, pliers holding the same white rectangle but now half the plastic is removed, final image is the PCB removed from the plastic rectangle, beneath this, from left to right a pile of plastic shredding, a coil of recycled black filament and a 3D printed boat.}
\end{figure*}

We demonstrated that a PCB can be recovered, as shown in Figure \ref{fig:deconstructionl} A-D: by heating up the prototype with a heat gun, cutting it open with a scalpel, and pulling out the PCB with pliers. The heat gun is used to bring the temperature of the plastic surrounding the PCB to the plastic's glass transition temperature of $60^{\circ}\text{C}$. Care should be taken not to overheat the part above the melting point of the solder at around $220^{\circ}\text{C}$, at which point components may be susceptible to being knocked off the PCB.

Prinjected material usually fuses to the top printed surface and remains adhered when peeled off, leaving the plated-through holes clear. However, if it remains lodged after extraction, it can be reheated and removed with a fine toothpick. 


We also show in Figure \ref{fig:deconstructionl} E-G that the conductive and non-conductive PLA can be shredded down and re-extruded to produce new filaments and printed into new objects. Using the mixed blend of conductive and non-conductive PLA from the shredded Printegrated Circuit housings and a 67:33 ratio of virgin to recovered material, we were able to extrude new filament with a Felfil filament extruder\footnote{Felfil filament extruder, \url{https://felfil.com/}, Accessed: 2025-04} and use it to print new objects. The resulting parts made from our recycled filament are non-conducting; this is due to the concentration of conductive filler being below the percolation threshold~\cite{taherian_electrical_2018}. However, as can be seen in Figure \ref{fig:deconstructionl}F and G, the recycled filament and resulting part appear black despite only a small amount of the black conductive material in the blend.

\section{Discussion}
Printegrated Circuits extends work around 3D printing interactive structures using available materials and tools to create self-contained devices so that they can be adopted and deployed by the community. In this section, reflecting on the work, we identify several areas that require further investigation to fully realise this goal.

\subsection{Design and Workflow Challenges}
The Printegrated Circuits process presents a distinct set of challenges to existing digital fabrication workflows. While the technique relies heavily on existing tools, the novel requirements introduced during design, slicing, and printing highlight the need for more flexible and extendable software. To support non-experts in designing new Printegrated Circuits, further work is required to encapsulate the domain-specific expertise traditionally gained from direct interactions with materials and machines. This would allow for extending existing tools to better support automation in this kind of workflow.

Specifically, better integration of aspects such as the Prinjection post-processing script and automated routing~\cite{schmitz_capricate_2015, savage_series_2014, wasserfall_topology-aware_2020} can enable non-experts to more easily harness these technologies. While the distribution and printing of existing designs, along with the mid-print insertion of PCBs, is conceivable for anyone with access to these tools, the complete design and fabrication process could be much better supported. Narrowing the access gap between experts and newcomers is critical and can be achieved by developing flexible and extendable tools that facilitate the exploration of new workflows~\cite{twigg-smith_dynamic_2023, tran_oleary_imprimer_2023, subbaraman_its_2025}.

\subsection{Material Limitations}
Printegrated Circuits overcomes some limitations posed by high resistance printable conductive material by using traditional electronics as material. Challenges still remain when considering modular embedded electronics, active sensing and actuation, or embedded energy storage, which all would require interconnects capable of carrying current between elements. While more conductive filaments such as Multi 3D's Electrifi copper filament~\cite{flowers_3d_2017} exist, they are still much more resistive than bulk metal, and have led researchers to explore the use of wet electroplating processes to improve it~\cite{hong_thermoformed_2021}. This kind of process comes with its own set of design and fabrication constraints, introducing new barriers to entry. Looking forward, we seek to explore embeddable electronic interconnects as well as functional electronics.

\subsection{Material-centred Free-form Personal Fabrication}
As additive manufacturing capabilities become increasingly accessible, designers and makers look to use these tools to fabricate more elements of their devices. DisplayFab~\cite{hanton_displayfab_2024} categorises interactive devices distinctly as input, output and control and distinguishes material-centric fabrication methods from component-based fabrication methods. It postulates the convergence of personally fabricated input techniques and the divergence of visual outputs or displays, with integration of control within material-centric methods being unattainable with current personal fabrication means. In Printegrated Circuits, we provide a hybridised approach offering a step beyond this limitation by spanning the integration of component-based control with material-centric 3D printing. We suggest that, using Printegrated Circuits as a baseline method, future work can explore the seamless integration of material-centric methods with component-based work. This harnesses the strengths of both systems to enable free-form multi-material structures with extended digital functionality within existing fabrication processes.


\subsection{Reversibility and Sustainability}
It is impossible to present the Printegrated Circuits technique without addressing the challenges around sustainability, specifically, the reversibility of the process in the form of disassembly for iteration and repair. We have demonstrated that the high-value general-purpose electronic elements can be recovered in a useful form and reused in new devices, and that the plastic can be partially recycled. 

Disassembly is destructive to any previously printed part of the object, and although the raw PLA material is derived from renewable resources, it does not break down freely in the environment. Other work overcomes some of these challenges with the use of screws~\cite{yan_solderlesspcb_2024} and magnets~\cite{schmitz_oh_2021}. Improved recycling techniques for multi-material prints~\cite{wen_enabling_2025} suggest that with improved ease of use and adequate access to fabrication and recycling tools, destructive recovery and full recycling of materials could be considered as a practical alternative to repair altogether. We also look to new and exciting work in extruded biomaterials~\cite{bell_biomaterial_2025} and consider how techniques to combine these with conductive material like charcoal could be used to build interactive devices with disposable and decomposable casings combined with re-usable embedded electronic elements.

\subsection{Future Printed Device Interaction Modalities}
In this paper, we presented a number of simple input modalities for Printegrated Circuits, however, existing research has covered many other 3D printed examples that could be augmented using our approach. We specifically look towards integrating output modalities within Printegrated Circuits, for example, embedding electroluminescent driver circuits within devices~\cite{hanton_protospray_2020} or integrating custom segmented electrophoretic `Eink' display film as demonstrated by Grosse-Puppendahl et al.~\cite{grosse-puppendahl_exploring_2016}. We are also interested in exploring techniques for combining flexible materials and electrostatic valves~\cite{gonzalez_layer_2022} to create actuated soft robotics.

We also imagine that the fully enclosed nature of these devices offers a wealth of opportunities beyond the creation of robust demonstrations.  Paired with wireless power transfer or energy harvesting techniques, and if necessary, a rechargeable battery or supercapacitor, fully dust-proof and water-resistant devices could be crafted for a range of interaction scenarios.

\section{Conclusion}

Personal fabrication research continues to democratise access to tools and processes, offering individuals new ways of turning their ideas into physical devices. In particular, 3D printed electronics has the potential to decentralise the production of interactive devices. However, challenges often emerge when taking these solutions beyond one-off demonstrations; limitations in robustness, materials and design tools present barriers to sustained deployment and scaling. This, in turn, limits sharing and adoption.

Printegrated Circuits lower these barriers by embedding PCBs into the 3D printing workflow with the help of new tooling. Our approach circumvents the limited current-carrying capacity of established conductive filaments by only using them for sensor structures and connections, keeping the remaining digital device interconnections, including power, within a traditional PCB. This technique enables the fabrication of interactive devices that are fully functional out-of-the-printer, supporting replicability and enabling small-batch production. Through a series of characterisations and demonstrations, we have presented that this approach is robust and ready to use within the existing digital fabrication workflow. We hope that Printegrated Circuits will inspire the fabrication of robust and fully integrated interactive devices with the tools we already have available to us as researchers, designers and makers, and we look forward to working with the community to extend the concept in future work.


\bibliographystyle{ACM-Reference-Format}
\bibliography{oliver_zotero}

\end{document}